\documentclass[a4paper,10pt]{scrartcl}

\usepackage{latexsym}
\usepackage{amsmath}
\usepackage{amsfonts}
\usepackage{amssymb}
\usepackage{graphicx}
\usepackage[english]{babel}
\usepackage{wasysym}
\usepackage{units}

\newcommand{\vs}[1]{\vspace{#1mm}}

\newcommand{\vsO}{\vspace{.1cm}\hfill\\}
\newcommand{\vsT}{\vspace{.2cm}\hfill\\}

\newcommand{\zo}{\zeta_{\scriptscriptstyle 0}}

\title{\Large GRAVITATIONAL INSTABILITY IN PRESENCE OF\\
BULK VISCOSITY: THE JEANS MASS AND\\
THE QUASI-ISOTROPIC SOLUTION}

\author{Nakia Carlevaro$^{\;a,\;b}$ and Giovanni Montani$^{\;a,\;b,\;c,\;d}$\vsT
\emph{\footnotesize $^a$ICRA -- International Center for Relativistic Astrophysics,}\vs{-2.5}\\
\emph{\footnotesize c/o Dep. of Physics - ``Sapienza'' Universit\`a di Roma}\\
\emph{\footnotesize $^b$Department of Physics - ``Sapienza'' Universit\`a di Roma, Piazza A. Moro, 5 (00185), Rome, Italy}\\
\emph{\footnotesize $^c$ENEA -- C.R. Frascati (Department F.P.N.), Via Enrico Fermi, 45 (00044), Frascati (Rome), Italy}\\
\emph{\footnotesize $^{d}$ICRANet -- C. C. Pescara, Piazzale della Repubblica, 10 (65100), Pescara, Italy}\vsO
{\footnotesize\ttfamily nakia.carlevaro@icra.it\quad montani@icra.it}
}
\date{}
\begin{document}
\maketitle

%
\hrule
\begin{abstract} \textbf{Abstract:} This paper focuses on the analysis of the gravitational instability in presence of bulk viscosity both in Newtonian regime and in the fully-relativistic approach. The standard Jeans Mechanism and the Quasi-Isotropic Solution are treated expressing the bulk-viscosity coefficient $\zeta$ as a power-law of the fluid energy density $\rho$, \emph{i.e.}, $\zeta=\zo\rho^{s}$. In the Newtonian regime, the perturbation evolution is founded to be damped by viscosity and the top-down mechanism of structure fragmentation is suppressed. The value of the Jeans Mass remains unchanged also in presence of viscosity. In the relativistic approach, we get a power-law solution existing only in correspondence to a restricted domain of $\zo$.
\end{abstract}
\hrule

\vspace{1cm}
\section{Characterization of viscosity}
\newcommand{\bv}{\textsc{Bv }}
The aim of this work is to study the dynamics of the gravitational instability, \emph{i.e.}, the evolution of scalar density perturbations, in presence of viscous effects. Viscosity can be divided in two different kinds: \emph{bulk viscosity} (\bv\!) $\zeta$ and \emph{shear viscosity} $\vartheta$. In this paper, we will concentrate on the analysis of isotropic (or almost quasi-isotropic) cosmological models. In this respect, we can safely neglect shear viscosity since there is no displacement of matter layers wrt each other (in the zeroth-order motion) and this kind of viscosity represents the energy dissipation due to this feature. Indeed, in presence of small inhomogeneities, such effects should be taken into account but we focus on the study of scalar density perturbations and volume changes of a given mass scale are essentially involved. We expect that the non-equilibrium dynamics of matter compression and rarefaction is more relevant than frictions, thus, we concentrate on \bv only since it outcomes from the difficulty for a thermodynamical system to follow the equilibrium configuration.

In this paper, we assume \bv as function of the state parameters of the fluid following the line of the fundamental analysis due to the \emph{Landau School} \cite{bk76}. In particular, we implement the so-called \emph{hydrodynamical description} of the fluid, \emph{i.e.}, an arbitrary state is consistently characterized by the particle-flow vector and the energy momentum tensor alone and viscosity is fixed by the macroscopic fluid parameters. In the homogeneous models $\zeta$ depends only on time and we assume it as a power-law of the density of the fluid $\rho$, \emph{i.e.}, 
\begin{align}
\zeta=\zo\,\rho^s\;,
\end{align}
with $\zo$, $s=conts.$

\section{Jeans Instability in Presence of Bulk Viscosity}
The study of the density-perturbation dynamics during the matter-dominated era can be consistently described using the Newtonian picture, as soon as sub-horizon-sized scales are treated. The fundamental result is the \emph{Jeans Mass} \cite{weinberg} ($M_J$), which is the threshold value for the fluctuation masses to condense, generating a real structure. If masses greater than $M_J$ are addressed, the density contrast ($\delta$) diverges (in time) giving rise to the gravitational collapse. 

In the following, we brief discuss how \bv affects such a scheme. The starting point is the Eulerian set of motion equations on which a perturbative theory is developed (the background is assumed static and uniform). If we introduce \bv effects in the first-order analysis, the density-contrast growth results to be dumped by viscosity, suppressing the structure formation \emph{without changing the threshold value of $M_J$}. In particular, the grater $\zo$ is, the slower $\delta$ diverges in time and, furthermore, viscosity generates a decreasing exponential regime and a damped oscillatory one in place of the standard pure oscillatory behavior \cite{nak-J}.

In this respect, we now analyze the \emph{top-down mechanism} of sub-structure formation, \emph{i.e.}, the comparison of one collapsing agglomerate with $M\gg M_J$ and an internal non-collapsing sub-structure with $M<M_J$. The sub-structure mass must be compared with a decreasing $M_J$ since the latter is inversely proportional to the collapsing agglomerate background mass. As soon as such value reaches the sub-structure mass, this begins to condense. In the standard Jeans Model, such mechanism is always allowed since the amplitude for perturbations characterized by $M<M_J$ remains constant in time but, the presence of decreasing $\delta$ in the viscous model, requires a discussion on the effective damping and on the efficacy of the top-down scheme. As shown in Figure 1, in correspondence of a very small viscosity coefficient (case A), we can show how the sub-structure survives in the oscillatory regime during the background collapse and the fragmentation occurs. On the other hand, if \bv is strong enough (case B), the damping becomes very strong and the sub-structure vanishes during the agglomerate evolution. Thus, the top-down mechanism results to be deeply unfavored.
\begin{figure}[h]\centering
\begin{minipage}[b]{5cm}
   \centering
   \includegraphics[width=\textwidth]{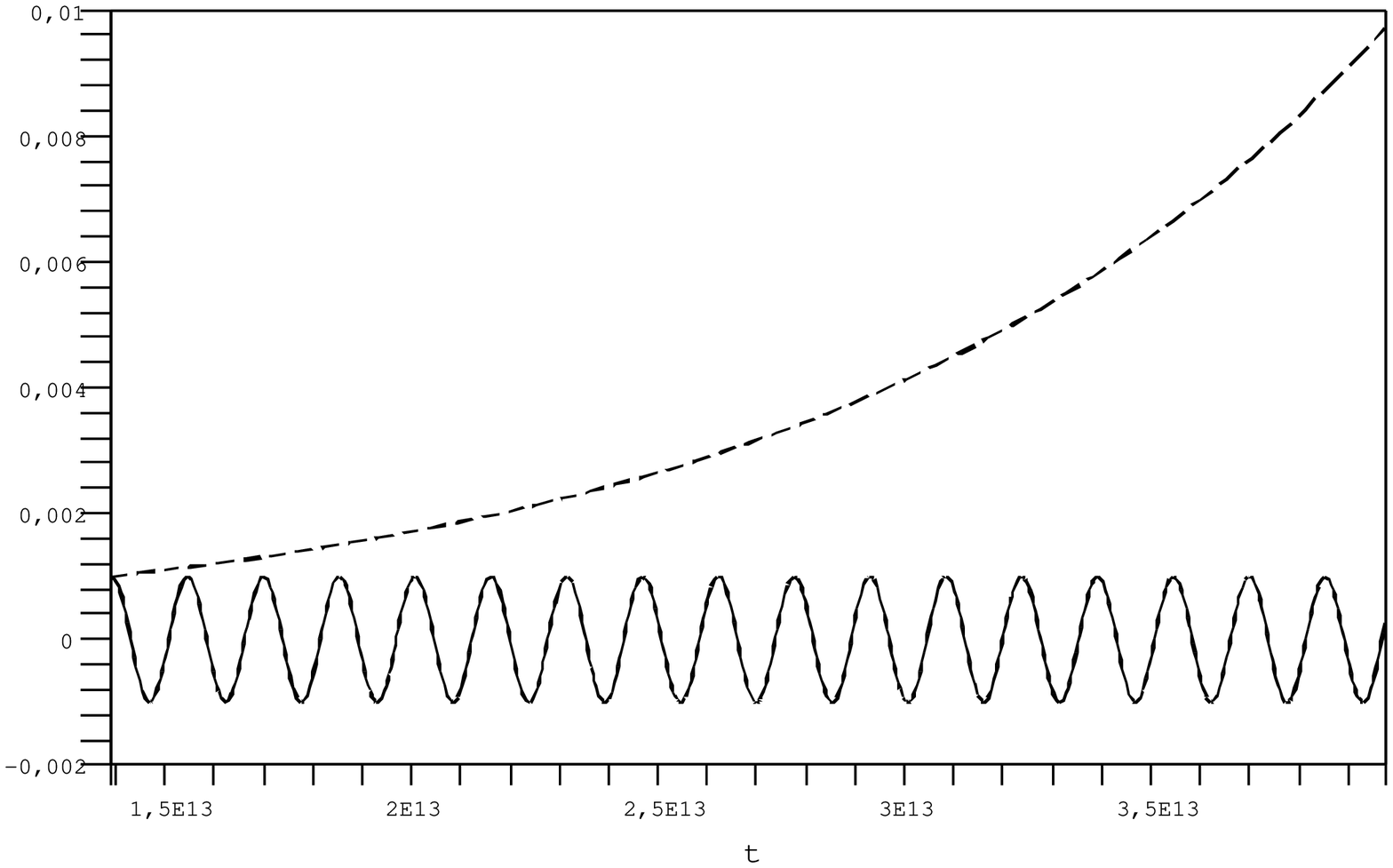}
   A
 \end{minipage}
 \begin{minipage}[b]{5cm}
  \centering
   \includegraphics[width=1.02\textwidth]{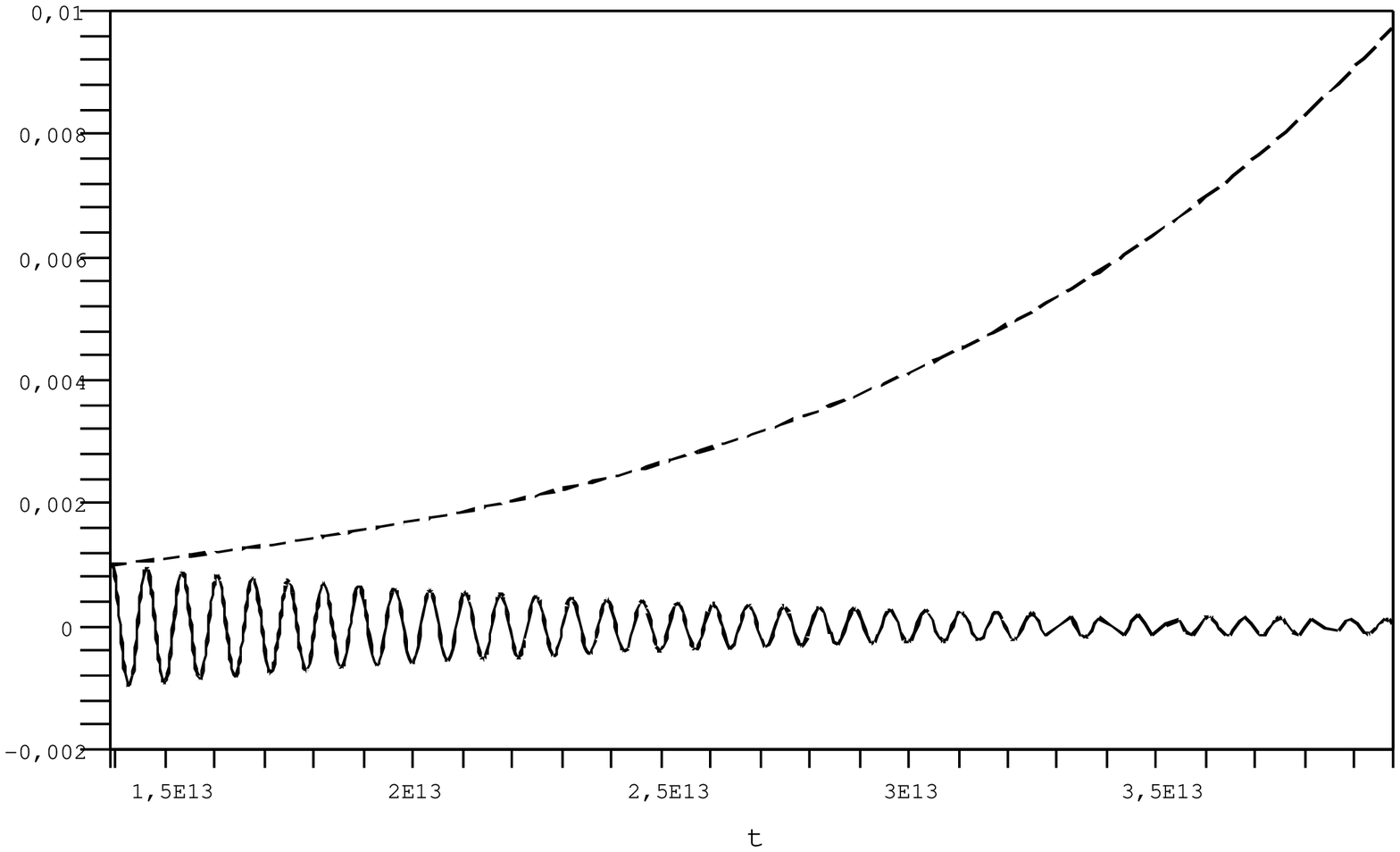}
   B
 \end{minipage}
\caption{\scriptsize{A -- $\zo=10^{-5}\,g/cm\,s$, galaxy $10^{12}\,M_\odot$: $\delta_G$ (dashed), sub-structure $10\,M_\odot$: $\delta_S$ (normal).
\qquad \qquad \quad \;
B -- $\zo=14\,g/cm\,s$, galaxy $10^{12}\,M_\odot$: $\delta_G$ (dashed), sub-structure $1\,M_\odot$: $\delta_S$ (normal).}}
\end{figure}

Moreover, if the static and uniform background solution is corrected for the expansion of the Universe \cite{nak-J}, a Jeans-like relation and a considerable damping of the density contrast growth can be found.

\section{Bulk-Viscosity Effects on the Quasi-Isotropic Solution}
In 1963 \cite{lk63}, E.M. Lifshitz and I.M. Khalatnikov firstly proposed the Quasi-Isotropic Solution (QIS) based on the idea that, as a function of time, the \emph{3}-metric $\gamma_{\alpha\beta}$ (where $\alpha$, $\beta$=1, 2, 3) is expandable in powers of $t$, \emph{i.e.}, a Taylor expansion is addressed:
\begin{equation}
\gamma_{\alpha\beta}=t^{x}\;{a}_{\alpha\beta}(x)+t^{y}\;{b}_{\alpha\beta}(x)\;,\qquad\quad
\gamma^{\alpha\beta}=t^{-x}\;{a}^{\alpha\beta}-t^{y-2x}\;{b}^{\alpha\beta}\;,
\end{equation}
where $x>0$ (space contraction) and $y>x$ (consistence of the perturbation scheme).

We now focus on the relevance of dealing with \bv properties of the cosmological fluid approaching the Big-Bang singularity \cite{nak-QI}. As far as we characterize the \bv coefficient like $\zeta=\zo\,\rho^s$, it is easy to realize that the choice $s=\nicefrac{1}{2}$ prevents dominating viscous effects. On the other hand, simple considerations indicate that the case $s<\nicefrac{1}{2}$ leads to negligible contributions of \bv in the asymptotic regime towards the Big-Bang. As a consequence, in studying the singularity physics, the most appropriate form of the power-law is $\zeta=\zo\,\sqrt{\rho}$ and our aim is to determine the conditions on the parameter $\zo$ (\emph{i.e.}, on the viscosity intensity) which allows for the existence of a QIS for the radiation dominated Universe.

To this purpose, we separate zeroth- and first-order terms into the 3-metric tensor and the whole analysis of Einstein equations follows this scheme of approximation. In the search for a self-consistent solution, we make use of the hydrodynamical equations in view of fixing the form of the energy density. Of course, in our solution the power-law for the leading \emph{3}-metric term is sensitive to $\zo$. As a result, by guaranteeing the consistence of the model, we find that the QIS exists if and only if $\zo$ remains smaller than a certain critical value, \emph{i.e.},
\begin{align}
\zo<\zo^{*}=\nicefrac{2}{3\sqrt{3}}\;,
\end{align}
(here $8\pi G=c=1$). In fact, for values of the viscous parameter greater than $\zo$, the perturbative expansion towards the singularity can not be addressed, since fluctuations would grow more rapidly than zeroth-order terms. 

It is worth noting that the study of the perturbation dynamics in a pure isotropic picture yields a very similar asymptotic behavior when viscous effects are taken into account \cite{nkm05}. The Friedmann scheme is preserved only if we deal with limited values of the viscosity parameter, obtaining the condition $\zo^{(iso)}<\zo^{*}/3$. This smaller constraint is physically motivated if we consider, \emph{as it is}, the Friedmann model a particular case of the QIS. Finally, analyzing $\delta$, we confirm and generalize the result obtained in \cite{nkm05} about the damping of density perturbations by the viscous correction.

\end{document}